\documentclass[twocolumn,aps,prb,groupedaddress]{revtex4-1}

\usepackage{graphicx,wrapfig,amsmath,upgreek,mathtools,amsfonts,physics,multirow,color}
\usepackage{graphicx}
\usepackage{color}
\usepackage{amsmath}
\usepackage{enumitem}
\usepackage{amssymb}
\usepackage{hyperref}

\newcommand{\ov}[1]{\bar{#1}}

\newcommand{\Z}{\mathbb{Z}}
\newcommand{\R}{\mathbb{R}}
\newcommand{\be}{\begin{equation}}
\newcommand{\ee}{\end{equation}}

\newcommand{\bea}{\begin{eqnarray}}
\newcommand{\eea}{\end{eqnarray}}

\newcommand{\p}{\partial}

\newcommand{\la}{\left\langle}
\newcommand{\ra}{\right\rangle}

\newcommand{\lp}{\left(}
\newcommand{\rp}{\right)}

\renewcommand{\vec}[1]{{\boldsymbol #1}}
\def\nn{\nonumber\\}

\newcommand{\addLL}[1]{\textcolor{blue}{#1}}
\newcommand{\addPL}[1]{\textcolor{magenta}{#1}}

\begin{document}
\title{Angular 
Superdiffusion and Directional Memory in Two-Dimensional  Electron Fluids} 
\author{Patrick Ledwith}
\altaffiliation{current address: Physics Department, Harvard University, Cambridge MA02138, USA}
\author{Haoyu Guo} 
\altaffiliation{current address: Physics Department, Harvard University, Cambridge MA02138, USA}
\author{Leonid Levitov} 
\affiliation{Department of Physics, Massachusetts Institute of Technology, 77 Massachusetts Avenue, Cambridge, Massachusetts 02139, USA}





\begin{abstract}
We demonstrate that 2D Fermi liquids can support peculiar  excitations that are not subject to Landau's $T^2$ dissipation. The long-lived excitations relax through correlated  angular dynamics involving ``lock-step'' angular displacements along the Fermi surface occurring in collinear two-particle collisions, a surprising behavior that is unique to 2D systems. We develop a microscopic picture of the non-Brownian random walk, describing the angular dynamics as anomalous diffusion (``superdiffusion'') on the Fermi surface. Strongly-correlated dynamics with directional memory, mediated by novel undamped excitations, dominates at moderately long times, pushing the onset of conventional hydrodynamics to abnormally large timescales. This exotic behavior can be directly probed by momentum-resolved tunneling 
techniques. 
\end{abstract}



\maketitle

\section{Hydrodynamics with directional memory}

\label{Sec1}
Electrons in two-dimensional (2D) materials 
can feature striking non-Fermi-liquid behaviors that are not found in ordinary solids. 
One celebrated example is Dirac electrons 
in graphene, 
where long-range electron-electron (ee) interactions invalidate the free-particle picture 
at charge neutrality\cite{gonzalez1994,gonzalez1999,vafek2007,song2007,katsnelson2007}. 
Strongly-correlated quasirelativistic Dirac fluids 
can exhibit a range of 
interesting quantum-critical and hydrodynamic 
behaviors \cite{
sheehy2007,fritz2008,muller2009, 
 lucas2016b, 
kashuba2018,lucas2018}. 

Here we 
argue that 2D Fermi gases with generic two-body interactions, e.g. such as those realized in graphene doped away from neutrality, also 
feature non-Fermi-liquid behavior, however of an entirely different kind. Namely, 
we predict dynamics with strong directional memory mediated by slow angular diffusion of 
excitations over the Fermi surface. 
Occurring at the timescales 
much longer than the conventional Fermi-liquid excitation lifetimes $\tau_* \sim 1/T^2$, such abnormally long-lived excitations define what 
may be appropriate to call a ``super-Fermi-liquid'' regime. 

Finite lifetimes of quasiparticles with characteristic temperature dependence $\tau_* \sim 1/T^2$ are an essential feature of Landau Fermi-liquids\cite{baym_pethick,coleman2015}. In three-dimensional (3D) systems, the timescale $\tau_* $ marks the onset of the hydrodynamic regime in which system memory of the microscopic state is fully erased. 
%
The 2D systems, however, have long been suspected to feature a behavior at $t>\tau_*$ that is considerably more complex and interesting than in 3D\cite{laikhtman_headon,gurzhi_headon,molenkamp_headon}. The unique aspect of 2D Fermi gases is that generic momentum-conserving particle collisions at a thermally blurred 2D Fermi surface are blocked by fermion exclusion, resulting in nontrivial constraints for the angles between momenta of colliding particles. The 
collinear processes singled out by these constraints, in particular their limited ability to provide angular relaxation, define a new dynamical regime for  two-dimensional Fermi systems\cite{ledwith2017,ledwith2019}. 
Understanding the nature and wide implications of this new ``super-Fermi-liquid'' behavior remains an outstanding challenge.

\begin{figure*}[!tp]
\centering
\includegraphics[width=0.8\textwidth]{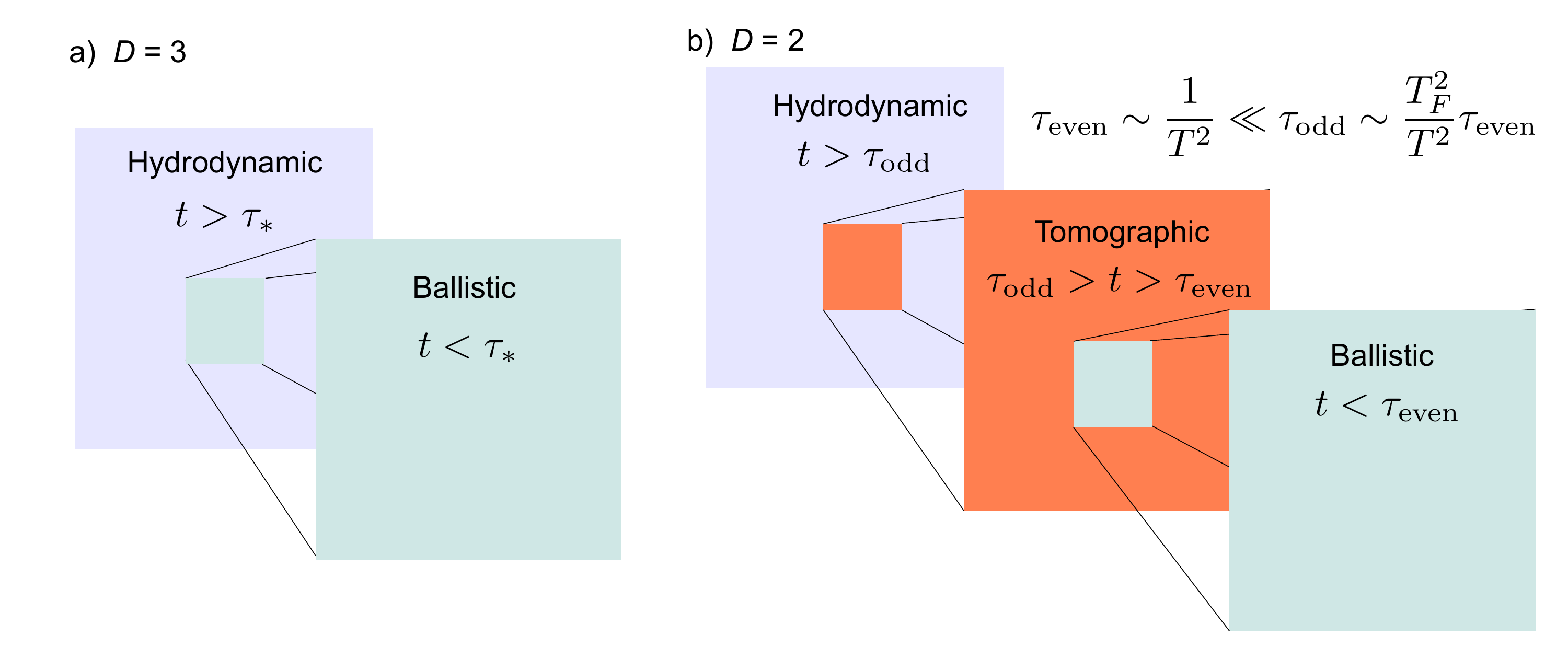}
\caption{Dynamical phases and time scales in Fermi liquids for different space dimensionalities. 
\textbf{a)} 
The conventional behavior in 3D: ballistic at short times and collective fluid-like at long times. 
\textbf{b)} 
The new, tomographic, regime arises in 2D between the conventional regimes, spanning a wide range of time scales, and featuring abnormally long-lived excitations and directional memory effects. The new behavior arises due to strongly-correlated angular displacements of colliding particles' momenta, and can be described as a superdiffusive non-Brownian random walk in angles, \eqref{eq:adiff}.
}
 \label{fig0}
\vspace{-5mm}
\end{figure*}

The goal of this work is to provide insight into some of these questions. 
As we will see, the dynamics at times $t>\tau_* $ represents a new transport regime which is distinct from conventional hydrodynamics. While many degrees of freedom relax at the ``normal'' time scales $t\sim\tau_* $, many other degrees of freedom remain unrelaxed and active at times 
$t\gg \tau_* $. Specifically, the dynamics is of a very different nature for the even-parity and odd-parity parts of momentum distribution, $f_{\pm}(\vec p)$, defined with respect to inversion $\vec p\to -\vec p$. Estimates show that the dynamical time scales for the odd-parity degrees of freedom are much longer than for the even-parity ones\cite{ledwith2019},
\be \label{eq:tau_odd_tau_even_a}
\tau_{\rm odd}\sim (T_F/T)^2\tau_{\rm even}
,\quad
\tau_{\rm even}\sim \tau_* 
.
\ee 
The new regime spans the wide range of time scales 
\be\label{eq:tau_odd_tau_even_b}
\tau_{\rm even} \ll t\ll \tau_{\rm odd}
,
\ee 
followed by a transition to the conventional hydrodynamic regime at the longest times, $t\sim \tau_{\rm odd}\gg \tau_*$. The hierarchy of time scales and the new regime are illustrated in Fig.\ref{fig0}. This hierarchy also translates, through the relation $\ell \sim v_F t$, into a new  hierarchy of spatial scales (see discussion in Sec.\ref{sec:summary}).

We will see that the dynamics of the odd-parity degrees of freedom features {\it directional memory} effects with long characteristic times, defining an interesting ``tomographic'' behavior.  The long timescales 
arise due to many repeated collisions with small angular stepsizes.  
We are therefore led to consider random walks on the Fermi surface arising due to small momentum changes in two-particle collisions. 
Processes of this type are routinely modeled as Fokker-Planck diffusion. 
The momentum-space Fokker-Planck picture is applicable to a wide variety of systems, ranging from collisions in plasmas and electron-phonon scattering in metals to radiation transport in cosmology\cite{LifshitzPitaevskii,sunyaev1970}. 

In our case, however, the random-walk dynamics on the Fermi surface 
comes with a twist that sets it apart from all previously known cases. 
We will show that this random walk is essentially non-Brownian, 
governed by special two-particle collisions, illustrated in Fig.\ref{fig:four_angles}, that are simultaneously {\it soft} and {\it head-on.} 
This leads to highly correlated angular displacements that are near-equal and opposite in each scattering process, enforced by kinematic constraints and fermion exclusion acting together. Such lock-step angular correlations 
produce a peculiar long-time angular dynamics of superdiffusive type: 
\be\label{eq:adiff}
\p_t \tilde f(\theta)=-D\p_\theta^4 \tilde f(\theta)
,\quad
\tilde f(\theta)=\frac{f(\theta)-f(\theta+\pi)}2
,
\ee
with $\theta$ the azimuthal angle on the Fermi surface and $\tilde f(\theta)$ the odd-parity part of momentum distribution. The angular (super)diffusion coefficient $D$ will be found to scale with temperature as $D\sim T^4/T_F^3$. The $T^4$ scaling, as well as the form of \eqref{eq:adiff}, are valid for not-too-small and not-too-large angular displacements, $\delta \theta\ll \theta\ll 2\pi$, with the UV cutoff 
$\delta\theta\sim \sqrt{T/T_F}$ and up to log corrections (see below). 

Anomalous diffusion, described by the square of Laplacian in \eqref{eq:adiff}, may seem to be at odds with the central limit theorem (CLT) that links random walks with diffusive transport in suitably defined configuration space\cite{van_kampen_1981}. The unique aspect of our problem, which invalidates this CLT-based intuition, has to do with nontrivial correlations of angular displacements in two-body collisions. The latter make the ``center of mass'' of the distribution in the angle variable 
conserved not just on average but individually in each collision process. This property suppresses the conventional {\it one-particle} diffusivity, generating instead the {\it two-particle} superdiffusive behavior. 

\begin{figure*}[!tb]
  \centering
  \includegraphics[width=\textwidth]{four_angles_combo_v6} 
  \caption{ \textbf{a)-c)} Different collision processes at thermally broadened 2D Fermi surface. Non-SH processes (a and c) each satisfy one of the conditions in \eqref{eq:headon_constraints}, the SH process (b) satisfies both conditions simultaneously. 
\textbf{d)} The dependence of scattering angles for ingoing and outgoing states $1$, $1'$, $2$, $2'$ and angular stepsizes $\Delta\theta$ vs.  momentum transfer $q$, illustrating that the SH processes dominate the stepsizes $\Delta\theta$. The angles $\theta_i$ are evaluated at a fixed energy transfer $\omega\sim T$, and measured from the direction perpendicular to $\vec q$. Stepsizes scale as $\sqrt{T}$ and $T$ for SH and non-SH processes (pictured next to the corresponding regions).  As discussed in the main text and supplement, the relevant angular displacements for hard head-on processes are $\Delta\theta_{1\bar 2}$ and $\Delta \theta_{\ov 2' 1'}$. Shown are the dependences found by resolving kinematic constraints, see \eqref{eq:thetas_vs_q}. The processes shown, which dominate angular stepsizes, correspond to small momentum transfers, $q\sim q_T\ll 2k_F$ (see \eqref{eq:q_T} and accompanying discussion). 
} 
  \label{fig:four_angles}
\end{figure*}

The educated reader would notice that the picture described above seems to contradict the analysis of excitation lifetimes based on the Green's function selfenergy  calculations, which for 2D Fermi liquids predicts log-enahnced decay rates that scale with temperature as $\gamma\sim T^2\log(T_F/T)$\cite{chaplik1971,hodges1971,bloom1975,giuliani1982,zheng1996,menashe1996,chubukov2003}. The selfenergy approach is therefore totally unaware of the existence of the long-lived excitations. 
This is because the selfenergy is most sensitive to the fastest decay pathways.  
If, as is often the case, there is a single timescale that characterizes decay for all low-energy excitations, the selfenergy is a signature of that timescale.
Yet, the single-timescale assumption is completely 
untrue 
in 2D, since the relatively slow angular diffusion creates a wide spectrum of lifetimes,  \eqref{eq:tau_odd_tau_even_b}. The 
selfenergy approach is therefore not well suited for such a situation.

The fact that slow angular dynamics pushes the onset of the conventional hydrodynamic behavior to very long times has wide implications for the ongoing quest for hydrodynamics in 2D electron fluids 
 \cite{andreev2011, 
forcella2014,tomadin2014,narozhny2015,principi2015,alekseev2016,narozhny2017,scaffidi2017,derek2018}, in which the directional memory effects have so far been ignored. We will comment on possible experiments that can probe angular dynamics in Sec.\ref{sec:summary}.
%
%

\section{Soft head-on processes and $\sqrt{T}$ stepsizes}
\label{Sec2}

To set the stage for analyzing superdiffusive behavior, we recall that kinematic constraints and fermion exclusion restrict momenta of particles colliding on a thermally broadened Fermi surface as \cite{laikhtman_headon,gurzhi_headon,molenkamp_headon} 
\be
\begin{aligned}\label{eq:headon_constraints}
&{\rm (i)}\quad \vec p_1=-\vec p_2, \quad \vec p_{1'}=-\vec p_{2'}
\\
&{\rm (ii)}\quad \vec p_1=\vec p_{1'}, \quad \vec p_2=\vec p_{2'}
\end{aligned}
\ee
where $1$, $2$ and $1'$, $2'$ label ingoing and outgoing states, respectively. These relations, which are true at leading order in $T\ll T_F$, and up to permutations of particles $1$ and $2$, imply that the even-parity and odd-parity degrees of freedom relax in very different ways. Namely, the even-parity degrees of freedom relax in a ``normal'' manner, through transferring particle distribution between distant points in momentum space in a single collision event, at a characteristic timescale $\tau_{\rm even}\sim 1/T^2$. In contrast, the odd-parity degrees of freedom remain unrelaxed at these timescales. 

As we will see, the odd-parity degrees of freedom are frozen only at leading order in $T\ll T_F$. Letting momenta go slightly off the Fermi surface (by an amount proportional to temperature) unfreezes these degrees of freedom and generates a random walk in momentum space. 
The resulting angular dynamics is dominated by the {\it soft head-on} (SH) processes for which the two conditions in \eqref{eq:headon_constraints}  are met simultaneously, such that all four momenta $\vec p_1$, $\vec p_2$, $\vec p_{1'}$, $\vec p_{2'}$ are near-collinear, as illustrated in Fig.\ref{fig:four_angles} b). As we will see, the SH processes define a non-Brownian random walk described as an anomalous diffusion with a square of Laplacian, \eqref{eq:adiff} with $D\sim T^4/T_F^3$. 

At the same time, the non-SH processes pictured in Fig.\ref{fig:four_angles} a) and c), taken alone, would lead to angular diffusion of a normal kind (see discussion in Sec.\ref{sec:superdiffusion}):
\begin{equation}
\frac{\p}{\p t}\delta f(\theta)=D'\partial_\theta^2 \delta f(\theta),\quad
D'\sim\frac{T^4}{T_F^3}
.
  \label{eq:diff}
\end{equation}
Comparing to the dynamics in \eqref{eq:adiff}, we see that the diffusion in \eqref{eq:diff} is relatively more slow than that in \eqref{eq:adiff} with the same diffusivity value, $D'=D$.
Indeed, in this case \eqref{eq:adiff} predicts a faster spreading 
than \eqref{eq:diff} for not-too-long times:
\be\label{eq:1/2>1/4}
\la\delta\theta^2\ra^{1/2}\approx (Dt)^{1/4}\gg (D't)^{1/2}
,\quad Dt\ll 1
,
\ee
representing nothing but the inequality $x^{1/4}\gg x^{1/2}$, $x\ll1$. 
\eqref{eq:diff} thus provides at most a subleading correction to \eqref{eq:adiff}. 

We note parenthetically that terms fourth order in gradients 
 usually arise in the theory of random walks as ``subdiffusion'' corrections to the diffusion equation. Indeed, gradient expansion in  $\xi\p_\theta$ with $\xi$ the 
 random walker stepsize would yield terms $\la \xi^2\ra\p_\theta^2$ and $\la \xi^4\ra\p_\theta^4$ with the prefactors of the same order. Then, because $\xi$ is of a microscopic scale, one would have $D\ll D'$. In our case, in contrast, for the SH processes the terms $\la \xi^2\ra\p_\theta^2$ vanish because of the lock-step correlations (see discussion in Sec.\ref{sec:superdiffusion}), and the only remaining nonzero second-order contribution is the one of non-SH processes. The latter, however, have much smaller stepsizes than the SH processes, resulting in $D\sim D'$. As a result, we have an interesting situation when the subdiffusion correction actually dominates. 
We will therefore use the name ``superdiffusion'' for \eqref{eq:adiff}.

The key difference between the SH and non-SH processes is in the size of angular displacements, which are order $\sqrt{T/T_F}$ for the SH processes and order $T/T_F$ for non-SH processes. These estimates can be obtained by expanding linearly in small deviations from the Fermi surface, such that particle energies are allowed to go off the Fermi level by an amount proportional to temperature.  The analysis involves resolving kinematic constraints for scattering angles and specializing to small energy transfers $\omega$ limited by $T\ll T_F$. For simplicity, we assume a general isotropic  carrier dispersion relation expanded in momentum deviations from the Fermi surface as
\begin{equation}
  \varepsilon(\vec p) = 
\varepsilon_F+\frac1{2m_*}(p^2-p_F^2), 
  \label{dispersion}
\end{equation}
with the terms higher-order in the expansion parameter $\eta=p^2-p_F^2$ omitted for brevity. 
The resulting scattering angles $\theta_1$, $\theta_{1'}$, $\theta_2$,  $\theta_{2'}$ dependence on momentum transfer $\vec q=\vec p_{1'}-\vec p_1=\vec p_2-\vec p_{2'}$ is obtained in Supplemental Information and is shown in Fig.\ref{fig:four_angles}. 

Angular relaxation is characterized by angular stepsizes $\Delta\theta$, which are marked in Fig.\ref{fig:four_angles}. These quantities have a nonmonotonic dependence on $q$ such that the maximal stepsizes 
are reached for SH processes, when $\vec q$ is {\it nearly perpendicular} to $\vec p_1$, $\vec p_{1'}$, $\vec p_2$, $\vec p_{2'}$. In this case, 
the momentum transfer is estimated as 
\be\label{eq:q_T}
q\sim q_T =\sqrt{2m_*T}
\ee 
with the energy transfer $\omega\sim T$ (see \eqref{energytrans} and accompanying discussion). Because $\vec q$ is a vector that connects $\vec p_1$ and $\vec p_{1'}$, we estimate the maximal angular displacement as
\be\label{eq:Delta_theta_11'}
\Delta\theta=\frac{q}{p_F}\sim \sqrt{\frac{T}{T_F}}
.
\ee  
In contrast, for non-SH process the momenta $\vec p_1$, $\vec p_{1'}$, $\vec p_2$, $\vec p_{2'}$ are not nearly perpendicular to $\vec q$. In this case, $v_Fq\sim\omega\sim T$ yields $\Delta \theta\sim T/T_F$.  For larger values of $q$ the collision becomes hard head-on.  As discussed in the next section and in the Supplemental Information, the relevant angular displacements in this case are $\Delta \theta_{1\bar 2} = \theta_1 - \theta_{\bar 2}$ and $\Delta \theta_{\bar 2' 1'} = \theta_{\bar2'} - \theta_{1'}$ where $\theta_{\bar 2} = \theta_2 +\pi$ and $\theta_{\bar2'} = \theta_{2'} + \pi$.  We will see that these stepsizes  also scale as $T/T_F$ for hard head-on collisions, but scale as $\sqrt{T/T_F}$ for SH collisions (as shown in Fig.\ref{fig:four_angles}).

This leads to two distinct types of angular steps: big ones for SH processes,  
and small ones for non-SH processes.  
Also, since $T\ll T_F$, typical 
$q$ values are quite small: $q/2k_F\sim \sqrt{T/T_F}\ll 1$. 
Accordingly, in Fig.\ref{fig:four_angles} we used \eqref{eq:thetas_vs_q} with $\omega=0.002T_F$. 

The scaling $\Delta\theta\sim T$ and $\Delta\theta\sim \sqrt{T}$ for angular steps can also be inferred from a purely geometric argument. One can note an interesting property of the collisions which are depicted in Fig.\ref{fig:stepsizes}: the tips of four momentum vectors form rectangles.  This rectangle property may be justified either by carrying out the above analysis further or, instead, by noting that the kinematics close to the Fermi level approximately follows a parabolic dispersion, \eqref{dispersion}. We may therefore boost to the center of mass frame, wherein the collisions become {\it perfectly} head-on. The scaling $T$ and $\sqrt{T}$ for angular steps then follows from inspecting the types of chords (momentum transfers) that can be drawn in an annulus of width $T$.

At this point, the reader may wonder whether the fine-tuned angles between momenta in SH collisions, which make the phase space for these processes small, will also reduce the SH collision rate. A detailed analysis shows that this is not the case\cite{ledwith2019}: instead, the two momentum delta functions are near-perfectly aligned and thus oversatisfied. This produces a well-known log enhancement of the collision rate, behaving as $\log(T_F/T) T^2/T_F$. For simplicity, this log enhancement will be ignored in subsequent analysis. 

\section{
Phase space for odd-parity degrees of freedom}
\label{sec:reflection}

  \begin{figure*}[!tb]
    \centering
      \includegraphics[width=\textwidth]{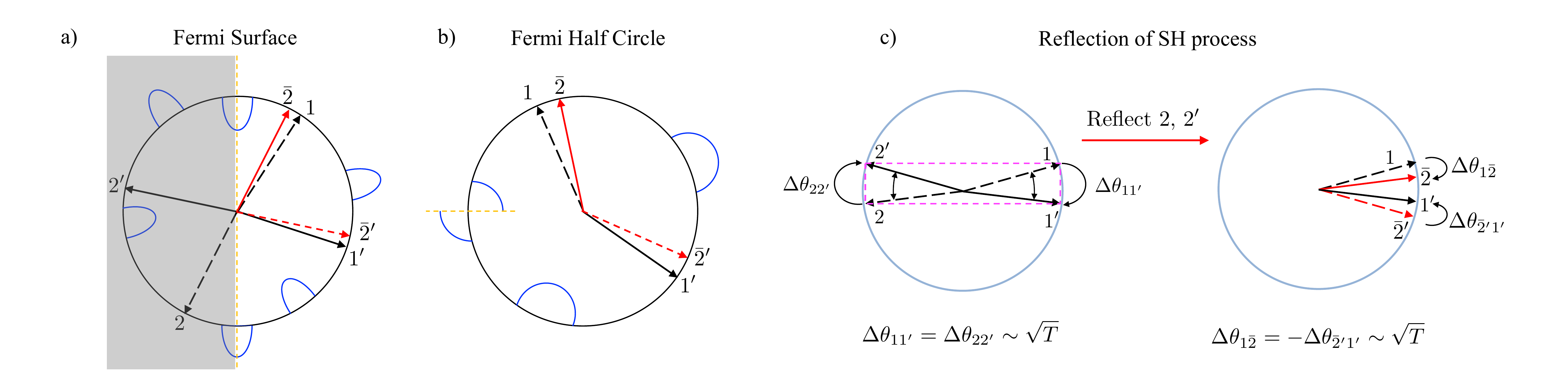} 
    \caption{
  \textbf{a)} Representing the odd-parity distribution on a half-circle with anti-periodic boundary conditions.  On the left is the Fermi surface with an odd-parity perturbation $\delta f$.  The black arrows labeled $1$ and $2$ (dashed lines) and $1'$ and $2'$ (solid lines) represent outgoing and ingoing momentum states in some collision, respectively.  The red arrows labeled $\bar 2$ and $\bar 2'$ are obtained by reversing $2$ and $2'$; 
solid and dashed lines are interchanged upon reversal, because, for an odd-parity distribution,
$\bar 2$ and $\bar 2'$ should be regarded as  {\it ingoing} and {\it outgoing} momentum states, respectively.      
In  \textbf{b)}
  the same objects are shown on an odd-parity configuration space, a half-circle formed by 
  gluing $\theta$ and $\theta + \pi$.   Antiperiodic boundary conditions are implemented at a branch cut joining $\theta=\pi/2$ and $\theta=-\pi/2$ (orange dashed line).
  \textbf{c)} Schematic of an SH collision and its transformation under reflection. 
The tips of ingoing and outgoing momenta $1$, $2$, $1'$, $2'$ form  rectangles (pink dashed line), as discussed in text.
  The angular steps $\Delta \theta_{11'}$ and $\Delta \theta_{22'}$ are equal and of order $\sqrt{T/T_F}$.  Replacing the 
states $2$ and $2'$ with the states $\bar 2$ and $\bar 2'$
gives a small-angle process which is local to a small section of the 
  half-circle.  The stepsizes $\Delta \theta_{1 \bar 2}$ and $\Delta \theta_{\bar{2}'1'}$ are also equal and of order $\sqrt{T/T_F}$.  The equality of stepsizes is a manifestation of 
lock-step correlations that give rise to non-Brownian random walks and angular superdiffusion.  
    } 
    \vspace{-5mm}
    \label{fig:adiff}
  \end{figure*}

Here we will introduce a Fermi half-circle configuration space, in which the dynamics of the odd-parity distributions
\be
  \delta f(\theta) = - \delta f(\theta + \pi).  
  \label{OPcond}
\ee
can be understood most clearly. 
The construction of half-circle configuration space is depicted in Fig.\ref{fig:adiff}b.  First, angles $\theta$ and $\theta + \pi$ are 
``glued together'' to form a circle 
of a two times smaller radius: 
\be
S^1/\Z_2 = \R\rm{P}^1 \simeq S^1
.
\ee  
In addition, anti-periodic boundary conditions are imposed on distribution functions, enforcing the odd-parity condition \eqref{OPcond}. 

To describe the change of  odd-parity distributions under collisions it is convenient to combine the odd-parity condition, \eqref{OPcond}, with symmetry between the ingoing and outgoing states, and introduce what we will refer to as ``reflection'' transformation. Indeed, for odd-parity distribution functions, adding a particle at $\theta$ 
is equivalent to removing a particle at $\theta + \pi$.  Thus, ingoing (outgoing) momenta at angle $\theta$ should be identified with outgoing (ingoing) momenta at angle $\theta + \pi$, respectively.  This is depicted in Fig.\ref{fig:adiff}a, with the ingoing and outgoing states 
$2$ and $2'$ and the corresponding reflected outgoing and ingoing states 
$\bar 2$ and $\bar 2'$ (red arrows). 

Applying this transformation to an SH 
collision pictured in Fig.\ref{fig:adiff}b, we reflect one of the ingoing states, $2$, and one of the outgoing states, $2'$, to obtain the outgoing state $\bar{2}$ and the ingoing state $\bar{2}'$.  
We may then view the processes as a two-particle hopping on a half-circle: $1 \to \bar 2$ and $\bar 2' \to 1'$, such that all states are positioned on the same small patch of the half-circle. The angular displacements after reflection, 
shown by arrows in Fig.\ref{fig:adiff}a, are then equal and opposite, \begin{equation}
  \Delta \theta_{1\bar{2}} = \theta_1 - \theta_{\bar{2}}, \quad \Delta \theta_{\bar{2}',1'} = \theta_{\bar{2}'} - \theta_{1'}
  ,
  \label{headonsteps}
\end{equation}
scaling with temperature as 
\begin{equation}
  \Delta \theta_{1\bar{2}} = -\Delta \theta_{\bar{2}'1'} \sim \sqrt{\frac{T}{T_F}}.
  \label{1dhoppingsteps}
\end{equation}
This defines a random walk process by which the odd-parity distribution changes,
which we will use below to 
analyze superdiffusion. 

The non-SH processes, which give a subleading contribution to angular dynamics, can be represented on the half-circle in a similar manner (see Figs.\ref{fig:stepsizes},\ref{fig:diff} and accompanying discussion).



\section{
Angular superdiffusion} 
\label{sec:superdiffusion}


Here we analyze the SH collisions 
and show that they lead to superdiffusive dynamics on the Fermi surface of the type in \eqref{eq:adiff}.
This behavior arises due to an interplay of two effects.  The first is that the angular stepsizes for SH processes are of order $\sqrt{T/T_F}$, as discussed above.
The second is the ``lock-step'' correlation in SH  scattering, by momentum conservation in the direction perpendicular to the collision axis; the lock-step property 
makes angular displacements in the half-circle configuration space equal and opposite, \eqref{1dhoppingsteps}.  Other types of collisions will be discussed below and shown to produce an ordinary diffusion, which is  subleading to \eqref{eq:adiff}. 

To gain insight into the anomalous properties of SH collisions, it is instructive to recall the general result for random walks based on the central-limit theorem (CLT). Namely, the long-time dynamics of a random walker is described by a diffusion equation with diffusivity 
\be \label{naivediff}
D_{\rm CLT} =r\la \xi^2\ra
,
\ee 
where $r$ is the rate at which the walker makes steps and $\la \xi^2\ra$ is the center of mass mean-square displacement in each step. For SH processes the  stepsizes are $\xi_\theta=\Delta\theta\sim\sqrt{T/T_F}$ and the collision rate is $r=cT^2/T_F$, where $c$ is an order-one constant. It is therefore tempting to conclude that a gradient expansion in 
$\xi_\theta \partial_\theta$ would yield a diffusion constant 
$D=r\la \xi_\theta^2\ra \sim T^3$.
However, the equality $\Delta \theta_{1\bar 2} = -\Delta \theta_{\bar 2' 1'}$, \eqref{1dhoppingsteps}, means that the lock-step-correlated angular steps do not result in the angular center of mass movement. This behavior, illustrated in Fig.\ref{fig:adiff} c), enforces $\xi=0$ and makes 
$D_{\rm CLT}$ vanish. 


From a microscopic perspective, we can explicitly perform a gradient expansion of a two-particle hopping $1, \, \, \bar{2}' \to \bar{2}, \, \,1'.$
  Accounting for the lock-step correlations, one generically obtains a CLT diffusion constant that vanishes if the stepsizes satisfy \eqref{1dhoppingsteps}: 
  \be\label{twohopdiff}
  D_{\rm CLT}= 
  r \la \left(  \Delta \theta_{1\bar 2} + \Delta \theta_{\bar 2'1'} \right)^2\ra =0
  ,
  \ee
as expected for the process in which the center of mass remains unchanged on each step,  $\xi =\theta_1 + \theta_{\bar 2'} -\theta_{\bar 2} - \theta_{1'}=\Delta \theta_{1\bar 2} + \Delta \theta_{\bar 2'1'}=0$. 
In this discussion we ignore the periodicity of the circle; this is 
legitimate because each SH process occurs on a small patch of odd-parity configuration space [see Fig.\ref{fig:adiff}a)], and so only after very long times will an initially peaked distribution ``become aware'' of the circle periodicity.

Since 
$D_{\rm CLT}$ in Eq.\ref{twohopdiff} vanishes, we need to go to higher orders in gradient expansion in stepsize. This can be done by defining 
a one-dimensional random walk model in which each step involves a correlated movement of pairs of different particles, as discussed in detail in Supplementary Material. At each step, particles move completely randomly but always in opposite directions to one another and in such a manner that their displacements are equal and opposite. In such a process, 
the center of mass is conserved in each movement of two particles, and thus the CLT diffusivity must vanish. 

Gradient expansion at fourth order in stepsize then indeed generates the superdiffusive behavior. The result in 
\eqref{eq:adiff} follows by going to fourth order in the angular-gradient expansion, necessary because the second-order CLT-based contribution vanishes, as discussed above.  On dimensional grounds, this predicts $\p_t \tilde f=-c\frac{T^2}{T_F}\left( \sqrt{\frac{T}{T_F}} \p_\theta \right)^4 \tilde f$, giving \eqref{eq:adiff}. 
Interestingly, the temperature dependence 
of $D'$ in \eqref{eq:diff} and $D$ in \eqref{eq:adiff} agree because the former is second order in the stepsize $T/T_F$ and the latter is fourth order in the stepsize $\sqrt{T/T_F}$.  Superdiffusion dominates, however, because the eigenvalues of $\partial_\theta^2$ are square integers, 
$\p_\theta^2 e^{im\theta}=-m^2e^{im\theta}$, 
and $m^4 \geq m^2$, which confirms our estimates in \eqref{eq:1/2>1/4}. 

One may see that processes satisfying \eqref{1dhoppingsteps} may not be described as ordinary diffusion cleanly from a macroscopic perspective.  If we start with a delta function distribution at $\theta = 0$, typical angular diffusion would cause the distribution to slowly spread out as a gaussian, with the variance of the gaussian growing with time.  We may interpret the growing variance as increased uncertainty in the center of mass.  But if our processes satisfy \eqref{1dhoppingsteps}, the center of mass is conserved \emph{for each process}, and hence its uncertainty must remain zero for all times.  The superdiffusive relaxation $\partial_t f = -  D\partial_\theta^4 f$ indeed obeys this requirement.  For a distribution 
that is initially 
peaked at zero, the time derivative of the variance of the center of mass vanishes:
$  \frac{\partial}{\partial t} \int_{-\infty}^{\infty} f(\theta,t) \theta^2  = -\int_{-\infty}^\infty D\frac{\p^4 f}{\p \theta^4} \theta^2=0
  $, with integration over $\theta$ carried out by parts.
Superdiffusion is therefore compatible with exact microscopic center of mass conservation.

\begin{figure*}[t]
\includegraphics[width=1\textwidth]{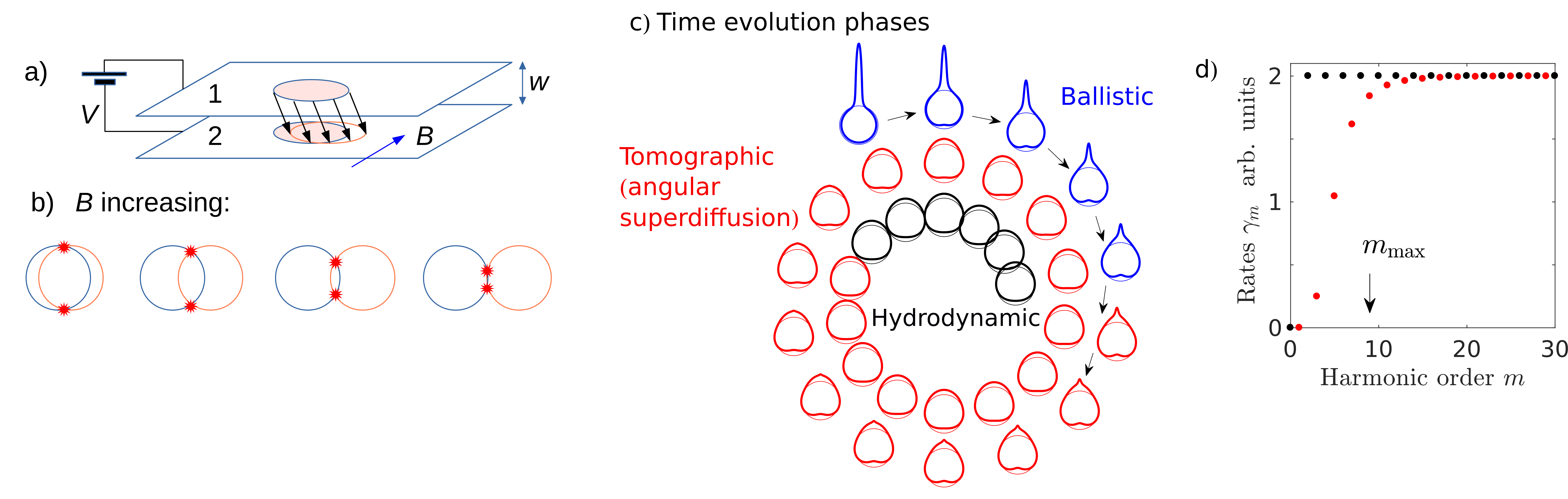}
\caption{a), b) Angle-resolved electron injection 
realized through momentum-conserving tunneling between parallel electron gases in the presence of a parallel magnetic field\cite{eisenstein_1991apl,eisenstein_1991prb,eisenstein_1995,burg_2018}. Momentum of a tunneling electron is shifted by $\Delta p=eBw$, where $w$ is the iterlayer spacing; as a result, tunneling between states at two Fermi surfaces is possible only near the hot spots (marked in red). By varying $B$ the hot spots can be brought close to each other, enabling angle-resolved injection and detection. c) Time evolution of a perturbed distribution 
 illustrating the multiscale dynamics arising due to the angular diffusion of the odd-parity distribution $\tilde f(\theta)$. 
The initial state is a delta-function---a bump on the Fermi surface---describing quasiparticle injected into the system. Three consecutive phases of time relaxation---ballistic, tomographic and hydrodynamic---are marked in blue, red and black, respectively. Even harmonics relax first, producing an odd-parity bump/antibump angular distribution. At longer times, the bump and antibump broaden, relaxing through angular (super)diffusion. Finally, only the $m=1$ harmonic survives, indicating transition to the hydrodynamic regime. The rates $\gamma_m$ used in the simulation are shown in panel d).
For illustration, we use a simple model:  identical values for even-$m$ rates, $\gamma_{m\ne 0}=\gamma_0$, and the dependence $\gamma_{m>1}=\gamma' m^4/(1+\gamma' m^4/\gamma_0)$ for odd-$m$ rates, with a small $\gamma'\ll\gamma_0$ and vanishing $\gamma_{m=1}$. The odd-$m$ rates behave as $\gamma'm^4$ at small $m$, saturating to a constant value $\gamma_0$ at large $m>m_{\rm max}=(\gamma_0/\gamma')^{1/4}$.
}
 \label{fig12}
\vspace{-5mm}
\end{figure*}

\section{
Experimental  implications} 
\label{sec:summary}

The slow modes that relax through angular diffusion can manifest themselves 
through interesting collective dynamics and nonlocal transport properties. 
Here we 
illustrate it for a well-studied experimental system that can serve as an analog of ARPES for 2D electrons, allowing the angular dynamics on the Fermi surface to be directly probed. It involves a pair of separately contacted 2D electron gases separated by a thin atomically-smooth barrier through which electrons can tunnel in a momentum conserving manner\cite{eisenstein_1991apl,eisenstein_1991prb,eisenstein_1995,burg_2018}. 
In the presence of a parallel magnetic field, momenta of  tunneling electrons shift by $\Delta p=eBw$ in the direction transverse to $\vec B$, where $w$ is the interlayer spacing. A small vertical voltage bias $eV\sim T\ll T_F$ induces interlayer tunneling 
that couples states at isolated points, where the $\Delta p$-displaced Fermi surfaces in the two planes intersect, as illustrated in Fig.\ref{fig12}. 

This setup allows electrons to be transferred between two Fermi surfaces in an angle-resolved manner, injecting electrons to (or, removing them from) two specific hot spots at the Fermi surfaces in each layer. The positions of the hot spots are tunable by varying the $B$ field strength. For Fermi surfaces of equal size, the two hot spots are separated by a field-tunable angle $2\arccos(B/B_0)$, $B_0=2\hbar k_F/ew$. Therefore, the angular distribution created through such a tunneling process has both an even-parity and an odd-parity part. The odd-parity part is small
at $B\ll B_0$, but grows with $B$ and reaches $50\%$ at $B\approx B_0$, when the hot spots approach each other. 
Such a system acts as a momentum-resolved electron injector 
with velocities of injected carriers 
collinear and tunable by varying the $\vec B$ field orientation and strength. 

After injection, the electron angular distribution will evolve in a complex manner, with several different timescales 
arising due to the even/odd parity asymmetry in relaxation rates and the superdiffusive dynamics, \eqref{eq:adiff}. As an illustration, here 
we consider a general time-dependent solution   
\be\label{eq:time_evolution}
f(\theta,t)=\sum_m f_m e^{im(\theta-\theta_0)} e^{-\gamma_m t}
\ee
where $f_m$ are Fourier harmonics of the initial 
perturbation at $t=0$, and $\theta_0$ is the injection direction. 

Different phases of time evolution, described by \eqref{eq:time_evolution}, are illustrated in Fig.\ref{fig12}. The distribution, initially localized at the angle $\theta=\theta_0$, first develops a bump/antibump structure of angular size which is small initially but begins to grow. Strikingly, the bump/antibump formation, which occurs on the single-collision timescales $t\sim \tau_*$, does not fully erase memory about the  initial state direction. The directional memory is lost at much longer times,  $t\gg \tau_*$, as the distribution continues spreading in angles over the entire Fermi surface. The long-time dynamics at $t>\tau_*$ can be viewed as an angular diffusion process through which the bump/antibump excitation gradually spreads over the entire Fermi surface. 
At very long times, only the $m=1$ harmonics remain, signaling that the system has eventually transitioned into the conventional hydrodynamic regime.

Detection of this complex time dynamics can be achieved by using a reverse of the process described above. Namely, applying a negative voltage bias will bring an electron back to the source layer provided that this electron is found at one of the tunneling hot spots at the Fermi surface. Through application of pairs of voltage pulses of opposite polarity with a controlled time separation, this geometry allows one to detect the dynamical phases of time evolution shown in Fig.\ref{fig12}. This approach, in combination with other transport measurements, can be used to map out the surprising new regime that lies between  the conventional ballistic and hydrodynamic regimes, in which the superdiffusive angular transport leads to peculiar directional memory effects, the lock-step dynamics and other surprising correlated behaviors. 

	\section*{Supplemental Information}
	\setcounter{section}{0}
	\renewcommand{\theequation}{S.\arabic{equation}}
	\setcounter{equation}{0}

\begin{figure}[!tb]
  \centering
  \includegraphics[width=0.47\textwidth]{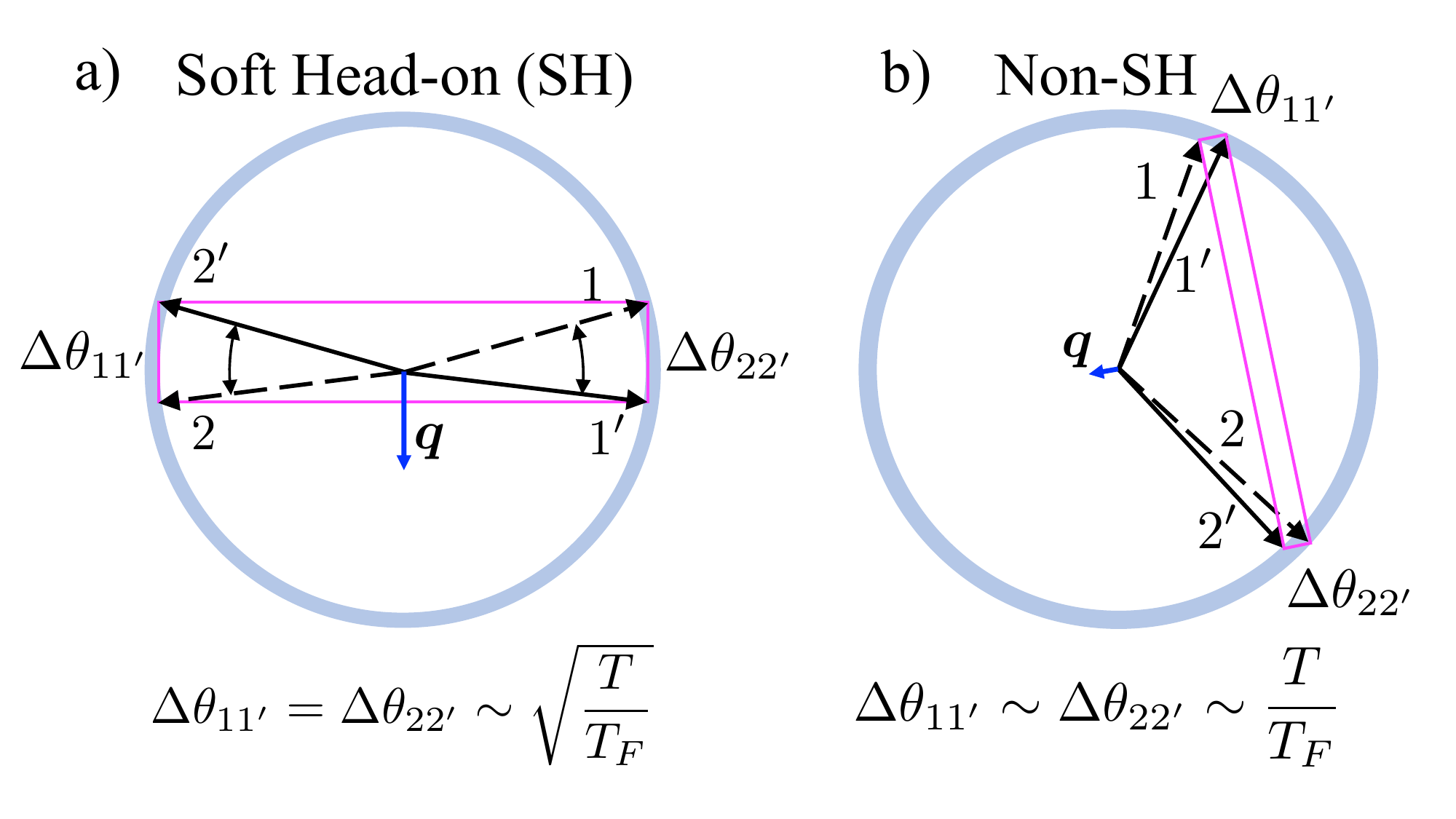}
  \caption{
Illustration of different scattering processes 
that give rise to different angular step sizes. 
Particle momenta are allowed to go off the Fermi surface by an amount proportional to temperature. Thermally blurred Fermi surface is shown in blue. Momentum states 
    $1$ and $2$ (dashed lines)
    are ingoing; 
    momentum states $1'$ and $2'$ (full lines) are outgoing. 
    \textbf{a)} 
    Soft head-on (SH) processes 
    that dominate angular relaxation.  
     In this arrangement, the momenta are all nearly collinear and nearly perpendicular to momentum transfer $\vec q$; as a result  the 
     scattering angles $\Delta\theta_{11'}$ and $\Delta\theta_{22'}$ are 
    of order $\sqrt{T/T_F}$.  
    \textbf{b)} 
    Generic large-angle non-SH processes that contribute to angular relaxation at subleading order in $T\ll T_F$. In this case, particle momenta are {\it not} nearly orthogonal to momentum transfer, thus the scattering angles are of order $T/T_F\ll \sqrt{T/T_F}$.
}
  \label{fig:stepsizes}
\end{figure}

\begin{figure}[!tb]
  \centering
  \includegraphics[width=0.47\textwidth]{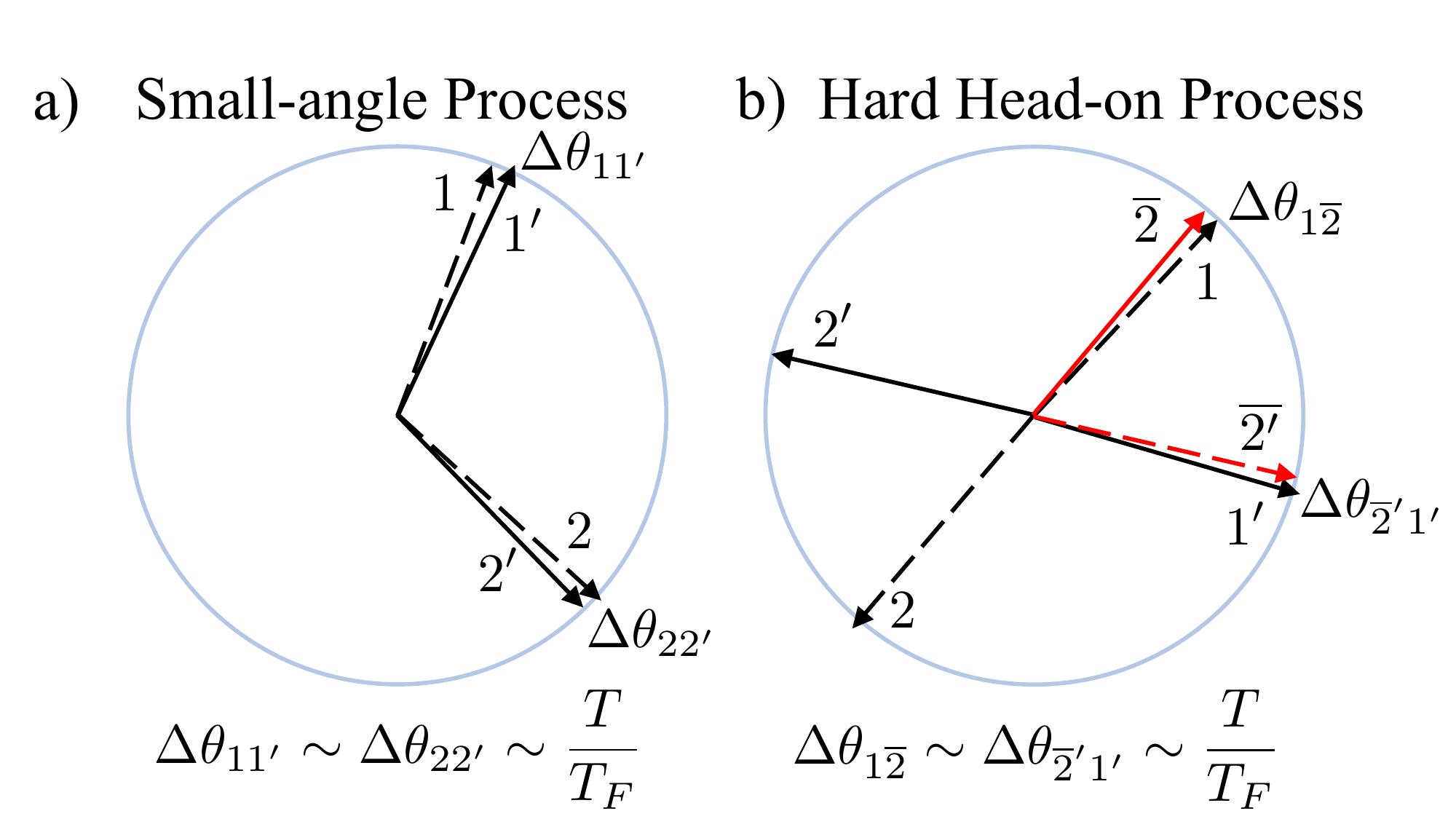}
\caption{Schematics of a) small-angle scattering and b) hard head-on scattering. 
States labeled $1$ and $2$ are incoming and $1'$ and $2'$ are outgoing.  For an odd parity distribution function, we may take $\theta_{2,2'} \to \theta_{2,2'} + \pi = \theta_{\bar{2}, \bar{2}'}$, where now $\bar{2}$ ($\bar{2}'$) labels an outgoing (ingoing) state.  Hard head-on processes are therefore small-angle processes on an odd-parity configuration space.  Both processes lead to angular diffusion \eqref{eq:diff}.
}
  \label{fig:diff}
\end{figure}

\section{Estimating angular stepsizes for different processes}
\label{sec:S1}

We consider general 
two-body collisions with ingoing momenta $\vec p_1$, $\vec p_2$ and outgoing momenta $\vec p_{1'}$, $\vec p_{2'}$. 
Owing to fermion exclusion, the energy transfer in 
such collisions is limited by temperature: 
\begin{equation}
  \omega = \frac{p_{1'}^2}{2m_*}-\frac{p_1^2}{2m_*} =\frac{p_{2}^2}{2m_*}-\frac{p_{2'}^2}{2m_*}\sim T.
  \label{energytrans}
\end{equation}
From now on it will be instructive to focus on the processes corresponding to small momentum transfer:
\begin{equation}
  |\vec q|\ll 2k_F
,\quad
\vec q = \vec p_{1'} - \vec p_1 
,
  \label{momentumtrans}
\end{equation}
where $k_F$ is the Fermi momentum. [The contributions of processes with $q$ of order $2k_F$ do not add anything new; they can be mapped on the $q\ll 2k_F$ processes by using a duality transformation that swaps the two outgoing states $1'$ and $2'$, as discussed in \cite{ledwith2019}.]
We may therefore 
write the energy transfer as 
\begin{equation}
    \omega  = 
  \frac{p_{1'}^2}{2m_*}-\frac{p_1^2}{2m_*} = v_1q\cos(\theta_1-\theta_q) +\frac{q^2}{2m_*}
  ,
  \label{etransexpand}
\end{equation}
where $v_1$ is the velocity of particle 1 and $\theta_q$ is the azimuthal angle of momentum transfer $\vec q$.

Using these relations we can estimate the angular displacements $\Delta \theta_{22'}$ and $\Delta \theta_{11'}$, pictured in Fig.\ref{fig:stepsizes},  as follows. 
First we consider generic values of $\theta_1 - \theta_q$, in which case the cosine is order one.  We therefore have $v_Fq \sim T$.  Because $\vec q$ is a vector that connects $\vec p_1$ and $\vec p_{1'}$, we obtain 
\begin{equation}
  \Delta \theta_{11'} \sim \frac{T}{T_F}
  \label{genericangles}
\end{equation}
and likewise for $\Delta \theta_{22'}$. 

However, a very different situation arises for $\theta_1 - \theta_q \approx \pm \pi/2$,  i.e. when $\vec q$ is nearly perpendicular to $\vec p_1$ and $\vec p_{1'}$. 
To estimate $q$ and $\Delta\theta_{11'}$ in this case, we rewrite the energy transfer \eqref{etransexpand} as follows:
\begin{equation}
    \omega  = 
  \frac{p_{1'}^2}{2m_*}-\frac{p_1^2}{2m_*} = v_{1'}q\cos(\theta_{1'}-\theta_q) -\frac{q^2}{2m_*}
  ,
  \label{etransexpand_1'}
\end{equation}
where $v_{1'}$ is particle $1'$ velocity. Subtracting this relation from \eqref{etransexpand} gives
\be
v_1\cos(\theta_1-\theta_q)-v_{1'}\cos(\theta_{1'}-\theta_q)=\frac{q}{m_*}
.
\ee
Introducing angles measured from the direction perpendicular to $\vec q$, 
$\tilde\theta_\alpha=\theta_\alpha-\theta_q+\pi/2$, $\alpha=1,1'$, 
and approximating $\cos(\theta_\alpha-\theta_q)\approx \tilde\theta_\alpha$, 
gives
\be\label{eq:theta1-theta1'}
v_F(\tilde \theta_1-\tilde \theta_{1'})=\frac{q}{m_*}
, 
\ee
where we approximated $v_{1'}$ and $v_1$ as $v_F$. Similarly, adding \eqref{etransexpand_1'} and \eqref{etransexpand} gives
\be
v_Fq(\tilde \theta_1+\tilde \theta_{1'})=2\omega
. 
\ee
Solving for $\tilde \theta_1$ and $\tilde \theta_{1'}$ gives values
\be\label{eq:thetas_vs_q}
\tilde \theta_1=\frac{\omega}{v_Fq}+\frac{q}{2m_*v_F}
,\quad
\tilde \theta_{1'}=\frac{\omega}{v_Fq}-\frac{q}{2m_*v_F}
.
\ee
The angles $\theta_{2}$, $\theta_{2'}$ can be obtained in a similar manner, they are given by \eqref{eq:thetas_vs_q} in which the energy transfer parameter $\omega$ is replaced by $-\omega$. In order to analyze angular stepsizes we use the angles
$\theta_{\bar 2}$, $\theta_{\bar 2'}$ which are obtained from $\theta_{2}$, $\theta_{2'}$ by a $\pi$ shift, as discussed in Sec.\ref{sec:reflection} and illustrated in Fig.\ref{fig:adiff}. The $q$ dependence for these angles is shown in Fig.\ref{fig:four_angles} of the main text. 

These dependences predict maximal angular displacements 
\be
\Delta\theta_{11'}=\tilde\theta_1-\tilde\theta_{1'}=\frac{q}{m_*v_F}
\ee 
for $q=q_\omega=\sqrt{2m_*\omega}$. Estimating $\omega\sim T$ and identifying $q_\omega$ with $q_T=\sqrt{2m_*T}$, as discussed in the main text, gives values
\be\label{eq:Delta_theta_11'_old}
\Delta\theta_{11'}\sim \sqrt{\frac{T}{T_F}}
,\quad
\Delta\theta_{22'}\sim \sqrt{\frac{T}{T_F}}
.
\ee  
Values $q< q_\omega=q_T$ yield smaller stepsizes $\Delta\theta$ that become very small as $q$ decreases, scaling linearly with $T$. 
At $q$ exceeding $q_T$, the 
displacements relevant for angular dynamics are found by performing a reflection transformation described in Sec.\ref{sec:reflection} and 
illustrated in Fig.\ref{fig:adiff}; these displacements are given by the quantities $\Delta\theta_{1\bar 2}$ and $\Delta\theta_{\bar 2'1'}$. At $q\gg q_T$, the stepsizes are also smaller than those in \eqref{eq:Delta_theta_11'}. Similar to the case $q\ll q_T$,  the 
quantities $\Delta\theta_{1\bar 2}$, $\Delta\theta_{\bar 2'1'}$ scale linearly with $T$ and become very small in the limit $q\gg q_T$,  corresponding to the 
processes shown in Fig.\ref{fig:four_angles} c). Because all momenta are nearly perpendicular to $\vec q$, they are all nearly collinear.  This type of small-angle collision is therefore nothing but an SH collision, 

Lastly, we note that the non-SH collisions 
lead to ordinary diffusion, \eqref{eq:diff}, with  a relatively small diffusion constant value, which makes them subleading to superdiffusion, \eqref{eq:adiff}.  Indeed, small-angle collisions with stepsizes $\Delta\theta\sim T/T_F$, as shown in Fig.\ref{fig:diff}a, lead to a nonzero diffusivity of value  
\be
  D' \sim \frac{T^2}{T_F}(\Delta \theta)^2 \sim \frac{T^4}{T_F^3}.
  \label{smallanglediff}
\ee
Here we used the central-limit theorem estimate for diffusivity $D'=r\la \xi^2\ra$, discussed in the main text. 
Generic hard head-on collisions also lead to the above result.  This can be seen by working with an odd-parity configuration space and reflecting particle states $2$ and $2'$ to $\bar 2$ and $\bar 2'$, as discussed in Sec.\ref{sec:reflection}. In this case, as
illustrated in Fig.\ref{fig:diff}b and discusses above, the stepsizes $\Delta \theta_{1\bar 2}$ and $\Delta \theta_{\bar 2'1'}$ 
scale as $T/T_F$, leading to \eqref{smallanglediff}. 

\section{A one-dimensional 
model for superdiffusion} 
\label{sec:S2}

Here we describe and solve 
a simple toy model that provides 
 an explicit derivation of superdiffusion for random walks that involve correlated lock-step movements of pairs of particles. 
In this model, particles hop on a line indexed by 
$-\infty<x<\infty$, a proxy for angle variable 
with the $2\pi$-periodicity ignored for simplicity. We assume a large bath of particles with density $n(x)$ changing in time due to movements of pairs of particles.  Namely, two particles separated by a distance $\ell$ hop in opposite directions, such that the one on the right hops $a$ to the right and the one on the left hops $b$ to the left.  We also include the time reversed processes where particles separated by a distance $\ell + a + b$ hop inward.  These processes occur with rate $\alpha n(x_1)n(x_2)$ where $x_{1}$ and $x_2$ are the positions of particle $1$ and $2$ before hopping (such that the distnce $|x_1 - x_2|$ is either $\ell$ or $\ell + a + b$).  For fermions, the hopping rates must also include 
additional factors of $(1-n)$ for the occupancy of final states. Here we drop these factors for conciseness; they may be included without changing the qualitative behavior of the model.
  
We now consider the influence 
of these processes on the evolution of the density $n(x)$.  We will see that the particles diffusively spread out as long as steps are unequal, $a \neq b$.  However, for equal steps, $a=b$, the diffusion constant vanishes due to center of mass conservation. In this case we obtain superdiffusive dynamics 
\begin{equation}
    \partial_t n = -D\partial_x^4 n
,
    \label{toysuperdiff}
  \end{equation}
with a nonzero superdiffusion constant $D$ 
which is a function of the model parameters.

The evolution of the probability distribution $\frac{d}{dt}n(x)$ receives eight contributions, one from each possible hopping involving the position $x$.  In particular, there are four positions involved in any hopping 
\be
x'-a-\ell-b,\quad x'-a-\ell,\quad x'-a,\quad x'
\ee
and the hopping can be inward or outward.  For the hopping where $x$ is all the way to the right, the contributions may be written as \begin{equation}
  \alpha n(x-a)n(x-a-\ell)-\alpha n(x)n(x-a-\ell-b),
  \label{farright}
\end{equation}
where for the first term the hopping is outward (initial positions $x-a$ and $x-a-\ell$) and for the second term the hopping is inward (initial positions $x$ and $x-a-\ell-b$).  The negative sign in the second term of \eqref{farright} arises because a particle is lost from position $x$, whereas in the first term a particle arrives at position $x$.  

The contributions from the other hoppings involving the position $x$ can be obtained by replacing $x \to x+a$, $x \to x+a+ \ell$ and $x \to x + a + \ell + b$, where for the replacements $x \to x+a$ and $x \to x+a + \ell$ we must multiply by an overall minus sign.  This is because for these positions $x$ is one of the inner positions and so outward (inward) hoppings result in particle loss (gain).  
We therefore have 
\begin{equation}
  \begin{aligned}
    \alpha^{-1} \frac{d}{dt}n(x) &= n(x-a)n(x-a-\ell) \\
    & - n(x)n(x-a-\ell-b)  \\
    & -(x\to x+a) -(x \to x+a+\ell) \\
    & + (x \to x+a+\ell+b).
\end{aligned}
  \label{baseeqn}
\end{equation}
We now linearize the above evolution equation close to an equilibrium density $n_0$.  Writing $n(x) = n_0 + \phi(x)$ and keeping terms linear in $\phi$ gives
\begin{equation}
  \begin{aligned}
\addLL{\gamma^{-1}}\frac{d}{dt}\phi(x) &= \phi(x-a)+\phi(x-a-\ell) \\
    & - \phi(x)-\phi(x-a-\ell-b)  \\
    & -(x\to x+a) -(x \to x+a+\ell) \\
    & + (x \to x+a+\ell+b).
\end{aligned}
  \label{linearize}
\end{equation}
where we introduced the notation $\gamma=\alpha n_0$.
Reordering the terms, we obtain 
\begin{equation}
  \begin{aligned}
\addLL{\gamma^{-1}}\frac{d}{dt}\phi(x) &= -\phi(x-a-\ell-b) - \phi(x+a+\ell+b)\\
    &+\phi(x+b+\ell)+\phi(x-b-\ell)\\
    &+\phi(x+a+\ell)+\phi(x-a-\ell) \\
    &+\phi(x+a)+\phi(x-a)\\
    &+\phi(x+b)+\phi(x-b) \\
    &-\phi(x-\ell)-\phi(x+\ell)-4\phi(x).
\end{aligned}
  \label{groupterms}
\end{equation}
Now, we perform coarse graining by carrying out gradient expansion. We
expand the terms on the right hand side as 
\begin{equation}
  \begin{aligned}
  \phi(x+\delta) &= \phi(x) + \delta\partial_x\phi(x) + \frac{1}{2}\delta^2\partial^2_x\phi(x)\\
  &+ \frac{1}{6}\delta^3\partial^3\phi(x)+\frac{1}{24}\partial^4\phi(x),
\end{aligned}
  \label{taylor}
\end{equation}
where $\delta$ represents the various deviations of the argument of $\phi$ from $x$ in \eqref{groupterms}.  We assume $a$, $b$ and $\ell$ are all of the same order.  The zeroth-order term vanishes because there are just as many terms with positive and negative signs.  The first-order and third-order terms cancel out by left/right symmetries.  The second-order terms give 
\begin{equation}
  \begin{aligned}
    \frac{d}{dt}\phi & = 
\gamma\big(-(a+b+\ell)^2+(b+\ell)^2+(a+\ell)^2 \\
    &+a^2+b^2-\ell^2\big)\partial_x^2\phi,\\
    &= \gamma(a-b)^2\partial_x^2 \phi,
    \end{aligned}
  \label{diffusion}
\end{equation}
which 
nothing but normal diffusion with diffusion constant $
\gamma (a-b)^2$.  

The diffusion constant vanishes for $a=b$, in which case we must 
proceed to a fourth-order expansion.  Collecting the fourth-order terms with $a=b$ we obtain 
\begin{equation}
  \begin{aligned}
  \frac{d}{dt}\phi & = \frac{
\gamma}{12}\left(-(2a+\ell)^4+2(a+\ell)^4+2a^4-\ell^4\right)\partial_x^4\phi, \\
  & = -
\gamma a^2(a+\ell)^2\partial_x^4\phi,
\end{aligned}
  \label{superdiff}
\end{equation}
and so the superdiffusion constant is $D = 
\gamma a^2(a+\ell)^2$ is nonzero even for $a=b$ as expected.  
Identifying $a$ with the $\sqrt{T}$ stepsizes in Eq.(20) 
of the main text, and estimating 
$\gamma$ as the base collision rate $\sim T^2/T_F$, 
we arrive at the dependence $D \sim T^4/T_F^3$ as anticipated in 
Eq.(3) of the main text.

\end{document}